\documentclass{moriond}

\usepackage{sidecap} %extra package I added

\def\be{\begin{equation}}
\def\ee{\end{equation}}
\def\bea{\begin{eqnarray}}
\def\eea{\end{eqnarray}}

\newcommand{\Photo}{}

\begin{document}
\vspace*{4cm}
\title{Using conditional GANs for convergence map reconstruction with uncertainties}

\author{Jessica Whitney, Tobías Liaudat, Matt Price, Matthijs Mars, Jason D. McEwen}

\address{Mullard Space Science Laboratory, University College London, \\
Holmbury St. Mary, Dorking, Surrey, RH5 6NT, England
}

\maketitle\abstracts{
Understanding the large-scale structure of the Universe and unravelling the mysteries of dark matter are fundamental challenges in contemporary cosmology. Reconstruction of the cosmological matter distribution from lensing observables, referred to as ‘mass-mapping’ is an important aspect of this quest. Mass-mapping is an ill-posed problem, meaning there is inherent uncertainty in any convergence map reconstruction. The demand for fast and efficient reconstruction techniques is rising as we prepare for upcoming surveys. We present a novel approach which utilises deep learning, in particular a conditional Generative Adversarial Network (cGAN), to approximate samples from a Bayesian posterior distribution, meaning they can be interpreted in a statistically robust manner. By combining data-driven priors with recent regularisation techniques, we introduce an approach that facilitates the swift generation of high-fidelity, mass maps. Furthermore, to validate the effectiveness of our approach, we train the model on mock COSMOS-style data, generated using Colombia Lensing’s $\kappa$TNG mock weak lensing suite. These preliminary results showcase compelling convergence map reconstructions and ongoing refinement efforts are underway to enhance the robustness of our method further.}

\section{Introduction}\label{sec:intro}

Incoming photons from distant galaxies travel along geodesics. Gravitational lensing refers to the phenomenon where these geodesics are perturbed due to intervening matter along the line of sight. Weak gravitational lensing is the regime where the geodesics are greater than one Einstein radius away from intervening matter at all times. Within this regime, lensing effects are small, and to observe the effect the aggregate over many galaxies must be taken.

To first order, weak lensing manifests in two ways: (i) the shear, $\gamma = \gamma_1 + i \gamma_2$: an anisotropic stretching of the source galaxy, (ii) the convergence, $\kappa$: isotropic magnification of the galaxy brightness. The convergence is unobservable directly, as there is no way to know the \textit{a priori} brightness of a galaxy, however, it contains a lot of useful information for cosmology. In particular, it traces the intervening matter distribution, and therefore is a way to probe dark matter. The shear, on the other hand, is an observable quantity. The problem of mass mapping is, therefore, how to calculate the convergence from the shear. This is an ill-posed inverse problem of the form 
$\gamma = \mathbf{A}\kappa + \mathbf{n}$,
where $\mathbf{A}$ is the lensing operator, and $\mathbf{n}$ represents systematic and instrumental noise. As we do not know the value of $\mathbf{n}$, calculating $\kappa$ is not a trivial task.

We propose using a conditional generative adversarial network (cGAN) for mass-mapping (for reviews of GANs see: Goodfellow et al. 2014; Goodfellow et al. 2020; Creswell et al. 2018). 
Deep learning methods have already proven to be promising when applied to mass-mapping  \cite{br}. Our aim is to provide a method with uncertainty quantification which has improved speed while maintaining high-quality reconstructions.

\section{Method and Preliminary Results}\label{sec:method}

We follow a similar approach to Bendel et al. (2024), including a regulariser to avoid the problem of mode collapse cGANs have previously been susceptible to. Our model consists of a uNET generator and a standard CNN discriminator (which outputs the estimated Wasserstein score). We used the $\kappa$TNG mock weak lensing suite \cite{ko} in conjunction with the Schrabback et al. (2010) COSMOS shape catalog to build a COSMOS-style dataset for use during training. We trained for 100 epochs on 4 A100 GPUs, which took approximately 6 hours.

% \begin{figure}
%        \centering
%        \includegraphics[width=0.5\textwidth]{our-gan.drawio.png}
%        \caption{An illustration of our conditional GAN framework.}
%        \label{fig:our-gan}
% \end{figure}

% \section{Preliminary Results}\label{sec:results}

\begin{figure}[h!]
       \centering
       \includegraphics[width=0.47\textwidth]{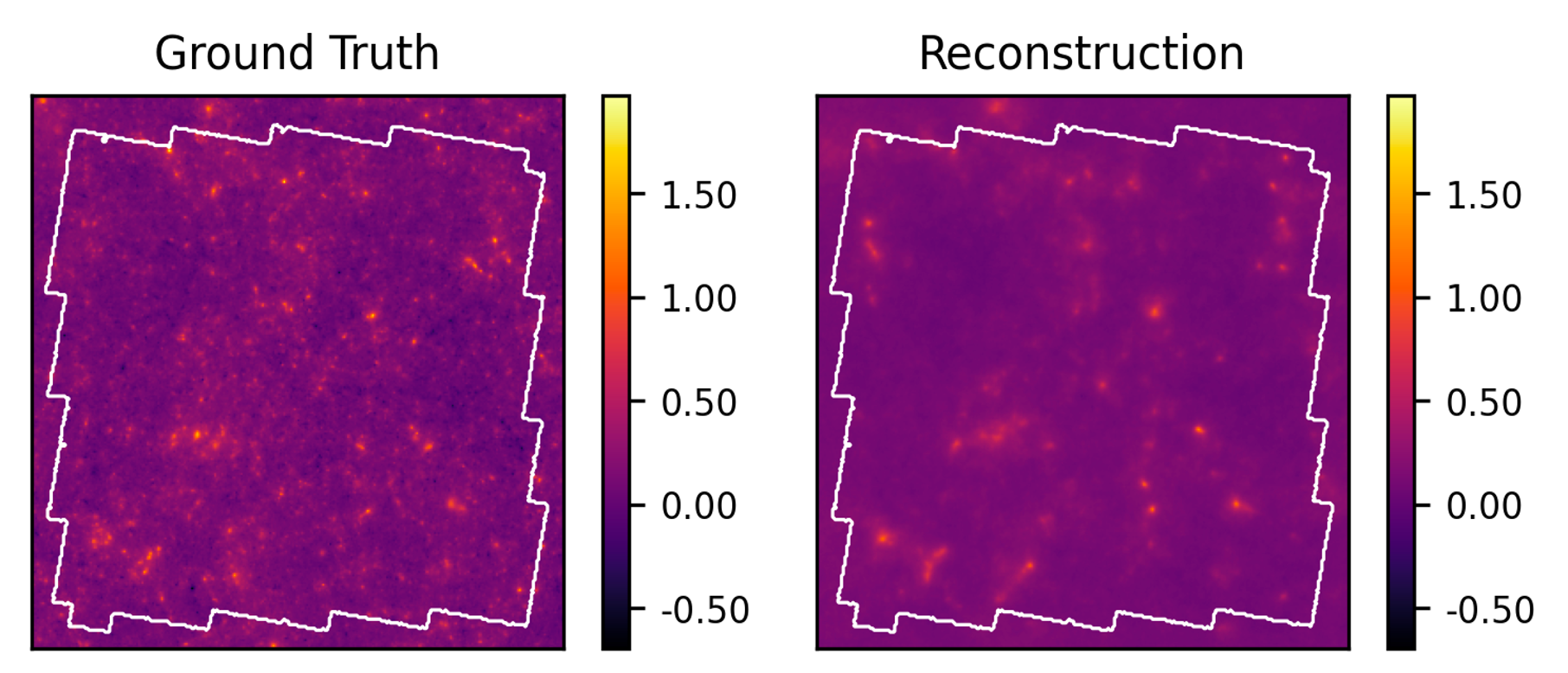}
       \includegraphics[width=0.47\textwidth]{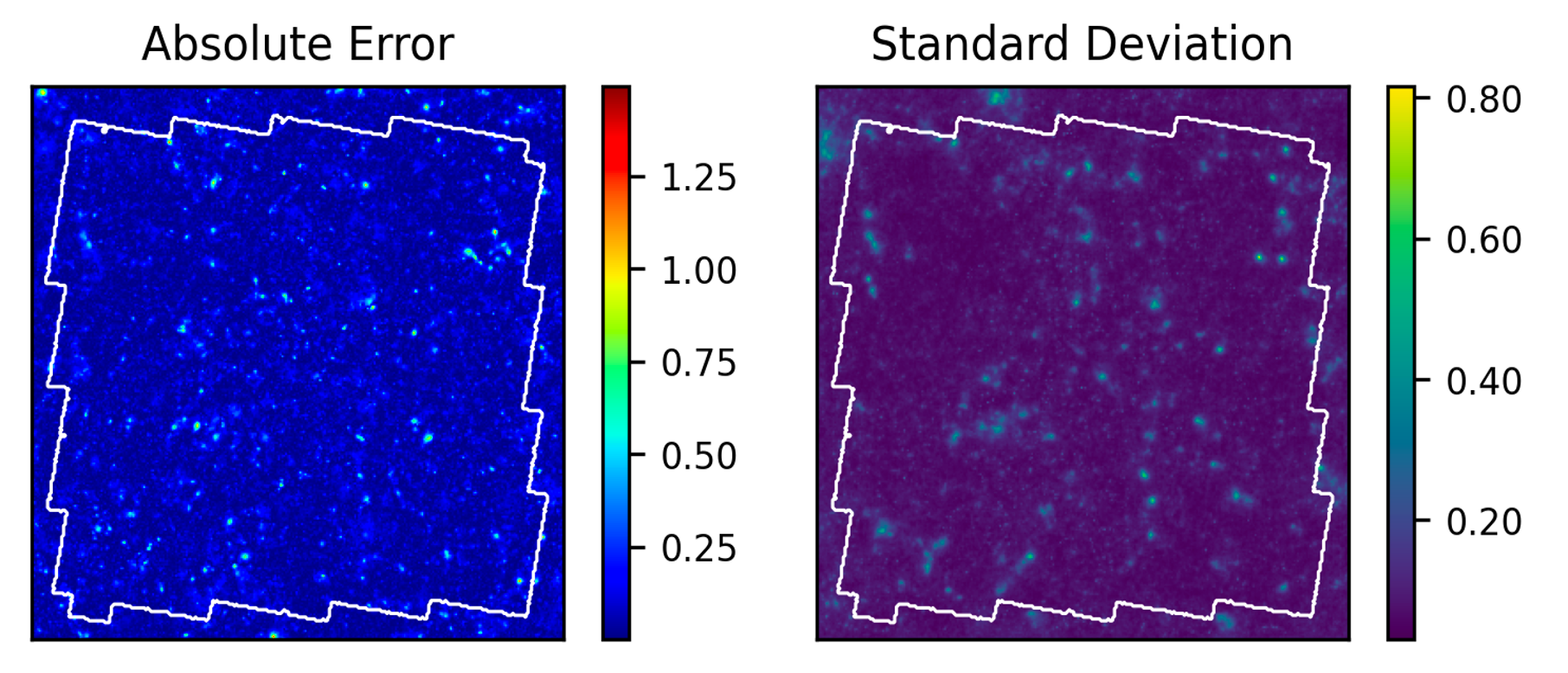}
       \caption{(Preliminary results). From left to right: a convergence map from our mock suite (ground truth); our (32-sample average) GAN reconstruction; the pixel-wise absolute error between the ground truth and reconstruction; and the standard deviation between the 32 posterior samples used to create the reconstruction. The standard deviation correlates with the true error, showing it is a good measure for the uncertainties associated with the reconstruction.}
       \label{fig:simulation-results}
\end{figure}

After training, when a given shear is passed through the generator it outputs a sample from the learned posterior distribution of the convergence. This process takes a matter of seconds. By passing the shear through $N$ times (we choose $N=32$), we may average over the $N$ posterior samples to create a convergence map reconstruction. For uncertainties, we take the standard deviation across the posterior samples used to build the reconstruction. Features which appear in more posterior samples represent features the cGAN is more certain about, these regions will have a smaller standard deviation, and vice versa, thus the standard deviation is a natural uncertainty metric. Figure \ref*{fig:simulation-results} highlights our preliminary results on the simulated data. It can be seen that the standard deviation correlates well with the pixel-wise absolute error between the truth and reconstruction, which again supports the reliability of the standard deviation as a measure of reconstruction uncertainty.

In conclusion, we have applied a novel cGAN framework to mass-mapping and validated our results on simulations, as seen in Figure \ref{fig:simulation-results}. Our method is data-driven, and can quickly generate high-fidelity reconstructions with uncertainty quantification. Full results will be available in (Whitney et al., in prep).
\section*{Acknowledgments}

This work was supported by EPSRC grant number EP/T517793/1, EPSRC grant number EP/W007673/1, and STFC grant number ST/W001136/1.

\end{document}